\documentclass[prb, showpacs, twocolumn,amsmath,amssymb]{revtex4}

\usepackage{graphicx}
\usepackage{dcolumn}
\usepackage{bm}
\pacs{74.43.-f}

\begin{document}


\title{
Electrical polarization of nuclear spins
in a breakdown regime of quantum Hall effect
}

\author{M. Kawamura}
	\email{minoru@iis.u-tokyo.ac.jp}
\author{H. Takahashi}
\author{K. Sugihara}
\author{S. Masubuchi}
\author{K. Hamaya}
\author{T. Machida}
	\email{tmachida@iis.u-tokyo.ac.jp}
\affiliation{Institute of Industrial Science,
 University of Tokyo, 4-6-1 Komaba, Meguro-ku, Tokyo 153-8505, Japan}

\date{\today}

\begin{abstract}
We have developed a  method for electrical polarization of 
nuclear spins in quantum Hall systems.
In a breakdown regime of odd-integer quantum Hall effect (QHE),
excitation of electrons  to the upper Landau
subband with opposite spin polarity dynamically polarizes
nuclear spins through  the hyperfine interaction.
The polarized nuclear spins in turn accelerate the QHE breakdown,
leading to hysteretic voltage-current characteristics 
of the quantum Hall conductor.
\end{abstract}

\maketitle


Control of nuclear spins in 
semiconductor has attracted considerable interests because
nuclear spin is one of the most promising 
elements for implementation of quantum bit.\cite{Kane1998}
Several techniques have been developed
for optical\cite{Salis2001, Sanada2006}
 and electrical\cite{Kane1992, Wald1994, Machida2002,
Machida2002_2, Machida2003, Kronmuller1999,
Hashimoto2002, Yusa2005, Ono2004}
control of nuclear spins.
In quantum Hall (QH) systems,
two kinds of approaches for all-electrical
manipulation of nuclear spins  have been demonstrated.\cite{Kane1992, 
Wald1994, Machida2002,
Machida2002_2, Machida2003, Kronmuller1999,
Hashimoto2002, Yusa2005}
One technique utilized spin-flip scattering of electrons
between  spin-resolved QH edge 
channels.\cite{Kane1992, Wald1994, Machida2002, Machida2002_2, Machida2003}
The flip of electron spin $\bm{S}$ flops  nuclear spin $\bm{I}$
through the hyperfine interaction,
$
{\cal H}_{\rm hyp} = A\bm{I}\cdot\bm{S} = 
A(I^{+}S^{-} + I^{-}S^{+})/2 + AI_{z}S_{z},
$
where $A$ is the hyperfine constant.
The nuclear spin polarization was detected
by measuring Hall resistance.
Another technique utilized 
domain structure with different spin configurations
in fractional QH systems.\cite{Kronmuller1999, Hashimoto2002, Yusa2005}
The spin-flip process of electrons traveling across the domain boundary
flops nuclear spins through the hyperfine interaction.

Nuclear spin polarization has been also utilized 
as a probe to investigate electron spin properties
in two-dimensional electron systems (2DESs),
which had not been accessed by standard magnetotransport measurements.
Indeed, the excitation of spin texture in a QH system\cite{Hashimoto2002}
and the low frequency spin fluctuations in closely separated
bilayer 2DESs\cite{Kumada2006}
were observed using  resistively detected nuclear spin relaxation.
Thus, development of a method for electrical
polarization and detection of nuclear spins 
will open a way to find spin-dependent  phenomena in QH systems.

In this letter, we demonstrate 
a  method for electrical polarization of  nuclear spins
using the breakdown of integer quantum Hall effect (QHE).
In a breakdown regime of odd-integer QHE,
electrons are excited to the upper Landau subband
with opposite spin polarity.
The spin-flip process of electrons dynamically polarizes
nuclear spins through the hyperfine interaction.
The polarized nuclear spins in turn
reduce the spin-splitting energy of Landau subbands,
accelerating the QHE breakdown.
The voltage-current characteristic curve is shifted
due to  the dynamical nuclear polarization (DNP).
The relevance of the DNP to the shift is
confirmed by  the detection of nuclear magnetic resonance (NMR).

\begin{figure}[b]
\includegraphics[width=8.0cm]{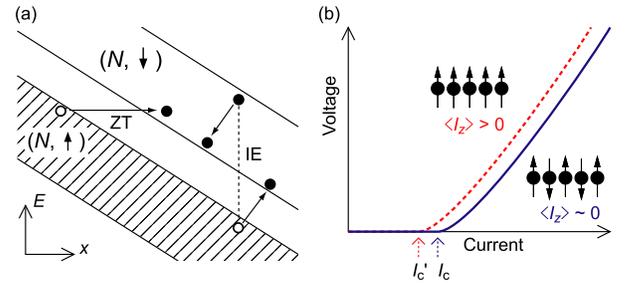}
\caption{\label{schematic}
(a) A schematic diagram of spin-split Landau subbands
in a breakdown regime of an odd-integer QHE.
Electrons are excited to the upper subband
through the Zener-type tunneling (ZT)\cite{Zener} and
the impact excitation (IE).\cite{Komiyama2000}
In the IE process,
electrons in the higher subband are
accelerated by the Hall electric field
to excite another electron in the lower subband
through the electron-electron scattering.
(b) A schematic representation of the expected
$V_{xx}$-$I$ curves at $\langle I_z \rangle \sim 0$
  (solid) and $\langle I_z \rangle > 0$ (dashed).
}
\end{figure}

\begin{figure}[t]
	\includegraphics[width=7.0cm]{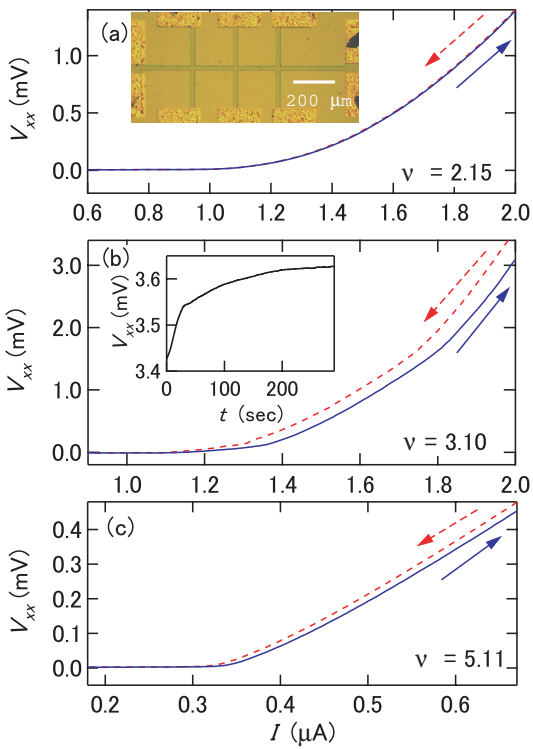}
	\caption{\label{hysteresis}
	$V_{xx}$-$I$  curves
	taken by sweeping current
	in positive (solid)
	and negative (dashed) directions at 
	(a) $\nu$ = 2.15 ($B$ = 5.15~T), 
	(b) $\nu$ = 3.10 ($B$ = 3.58~T),
	and (c) $\nu$ = 5.11 ($B$ = 2.17~T).
	Inset of (a):  A micrograph of the Hall-bar device.
	Inset of (b): Time evolution of $V_{xx}$
	after the current is changed from $I$ = 0.0~$\mu$A
	to 2.0~$\mu$A at $t$ = 0.
	}
\end{figure}

We propose a concept for electrical polarization of nuclear spins
 in an odd-integer QHE regime,
where the Fermi energy resides in the energy gap
of spin-split Landau subbands.
In this condition, the lower Landau subband ($N$, $\uparrow$) is fully occupied
with up-spin electrons, while the higher down-spin subband 
($N$, $\downarrow$) is empty.
When a current is transmitted through the conductor,
the Landau subbands are tilted due to the Hall electric field
as schematically shown in Fig.~\ref{schematic}(a).
When the current is increased above a critical current $I_{\rm c}$,
electrons in the lower Landau subband (up-spin)
are excited to the higher empty subband (down-spin),
giving rise to an abrupt increase of longitudinal
voltage $V_{xx}$.
This phenomenon is referred to as the QHE breakdown.\cite{Ebert1983, Cage1983}
Possible excitation processes of electrons include  
the Zener-type tunneling (ZT)\cite{Zener} 
and the impact excitation (IE)\cite{Komiyama2000}
[Fig.~\ref{schematic}(a)].
Though the mechanism of the QHE breakdown has been still under debate,
it is obvious that the excitation processes 
are accompanied by up-to-down spin flips of electrons.
Accordingly, 
we expect that the QHE breakdown 
can be utilized to polarize nuclear spins,
i.e.
the up-to-down spin flips of electrons
dynamically polarize the nuclear spins
along  the external magnetic field $B$ ($\langle I_z \rangle > 0$)
through the counter spin flops of nuclear spins
in the hyperfine interaction\cite{energyconservation}.

When the nuclear spins are polarized in the QHE breakdown regime,
the polarized nuclear spins ($\langle I_z \rangle > 0$)
reduce the spin-splitting energy
$
E_{\rm S} = |{\rm g}|\mu_{\rm B}B - A\langle I_z \rangle,
$
where g is  the g factor of electrons (= $-$0.44 in GaAs) and
$\mu_{\rm B}$ is the Bohr magneton.
Since the odd-integer QHE is stabilized by $E_{\rm S}$,
the reduction of $E_{\rm S}$ 
is expected to accelerate the QHE breakdown,
leading to the shift of voltage-current ($V_{xx}$-$I$) curves
toward the smaller current side as shown in Fig.~\ref{schematic}(b).
Thus, the $V_{xx}$-$I$ curve
is expected to show hysteresis
depending on the sweep direction of the current.
Nachtwei {\it et al.} observed hysteretic $V_{xx}$-$I$ curves
in InGaAs/InAlAs systems, 
but they excluded  the relevance of nuclear spins 
and  interpreted the hysteresis 
in terms of quantum Hall ferromagnet.\cite{Nachtwei2000}
Song and Omling also reported hysteretic magnetotransport 
in the regime close to the QHE breakdown.\cite{Song2000}
They found an unexpected huge differential resistance peak
with a very slow relaxation time and suggested
the influence of nuclear spins on it.
However, the relationship between the nuclear spins and the QHE breakdown
has been unclear.

A Hall-bar device with a channel width of 20~$\mu$m
was fabricated by photolithography from a wafer of
GaAs/AlGaAs single
 heterostructure\cite{wafer}[inset of Fig.~\ref{hysteresis}(a)].
The mobility and  sheet carrier density of the 
2DES are $\mu$ = 60~m$^{2}$/Vs
and $n$ = 2.7 $\times 10^{15}$~m$^{-2}$, respectively.
The longitudinal voltage $V_{xx}$ was measured
by a standard dc four-terminal method
in a dilution refrigerator with 
a base temperature of 20~mK.
A single-turn coil around the device was used to irradiate
radio-frequency (rf) magnetic fields.


\begin{figure}[t]
	\includegraphics[width=7.5cm]{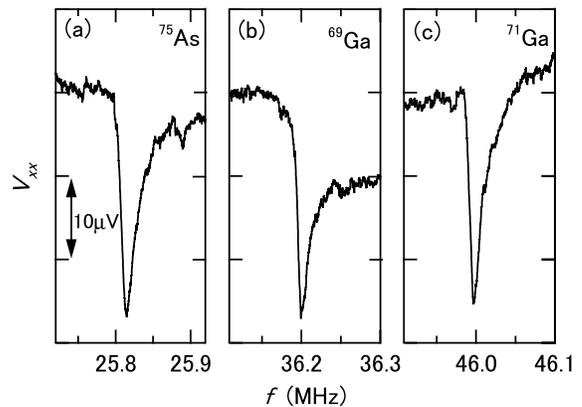}
	\caption{\label{NMR} 
	NMR spectra for (a)$^{75}$As, (b)$^{69}$Ga, and (c)$^{71}$Ga
	detected by measuring $V_{xx}$. 
	The sweep rate of rf magnetic fields is 13 kHz/min.
	}
\end{figure}

Voltage-current curves in QHE regimes were taken by
sweeping the current between $-$2.5 $\mu$A and 2.5 $\mu$A
at various Landau-subband filling factors $\nu$.
Figures~\ref{hysteresis}(a)-\ref{hysteresis}(c) show
the $V_{xx}$-$I$ curves in QHE regimes 
at $\nu$ = 2.15 ($B$ = 5.15~T),
$\nu$ = 3.10 ($B$ = 3.58~T),
and $\nu$ = 5.11 ($B$ = 2.17~T).
The solid and dashed curves are respectively
obtained by sweeping the current at a rate of 0.012 $\mu$A/s
in positive and negative directions.
The value of  $V_{xx}$ starts
to increase at $I$ = 1.0 $\mu$A, 1.1 $\mu$A, and 0.33 $\mu$A
for $\nu$ = 2.15, 3.10, and 5.11, respectively.
At  $\nu$ = 3.10 and 5.11
[Figs.~\ref{hysteresis}(b) and \ref{hysteresis}(c)],
the shift of 
the down-sweep curves toward the smaller current side
is observed, 
while no shift is found
at $\nu$ = 2.15 [Fig.~\ref{hysteresis}(a)].
The observed shift of the  $V_{xx}$-$I$  curves
is consistent with our expectation [Fig.~\ref{schematic}(b)].

The inset of Fig.~\ref{hysteresis}(b) shows
time evolution of $V_{xx}$ at $\nu$ = 3.10 
after the current is changed from $I$ = 0.0~$\mu$A to
2.0~$\mu$A.
The value of  $V_{xx}$ increases slowly 
with a relaxation time over 300 s,\cite{negativeI}
which is a typical time scale for the nuclear spin relaxation.\cite{Kane1992,
Wald1994,Machida2002,Kronmuller1999,Hashimoto2002,Yusa2005,
Machida2002_2, Machida2003}
The increase in $V_{xx}$ indicates 
the acceleration of  the QHE breakdown due to 
the reduction of the spin-splitting energy $E_{\rm S}$.

The relevance of DNP to the shift of the $V_{xx}$-$I$ curves is 
unambiguously confirmed by the NMR measurements described below.
A rf magnetic field parallel to the 2DES is applied
after $V_{xx}$ is completely saturated
at $I$ = 2.0~$\mu$A.
The value of $V_{xx}$ decreases
at the NMR frequencies of $^{75}$As, $^{69}$Ga, and $^{71}$Ga
as shown in Figs.~\ref{NMR}(a)-\ref{NMR}(c).
The detection of  NMR  shows that
the shift of the $V_{xx}$-$I$ curves are caused by the DNP
and that the nuclear spins are polarized 
in the   QHE breakdown regime.

\begin{figure}[t]
	\includegraphics[width=7.0cm]{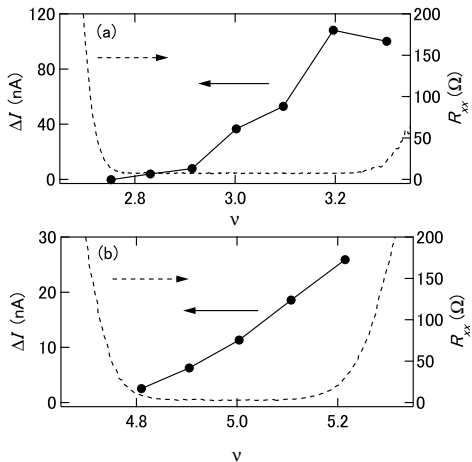}
	\caption{\label{fielddep}
	Shift of $V_{xx}$-$I$  curves
	between up- and down-sweeps at $V_{xx}$ = 2~mV
	as a function of the Landau-subband filling factor
	in the QHE plateau regions of (a) $\nu$  = 3 and (b) $\nu$ = 5.
	The longitudinal resistance $R_{xx}$ is plotted together
	by the dashed curves.
	}
\end{figure}

The shift of the $V_{xx}$-$I$  curves is
prominent in the odd-integer QHE plateaus
of $\nu$ = 3 and 5 [Figs.~\ref{hysteresis}(b) and \ref{hysteresis}(c)],
while it is almost absent 
in the even-integer QHE plateaus 
of $\nu$ = 2, 4, and 6  [Fig.~\ref{hysteresis}(a)], 
where the cyclotron energy $\hbar\omega_{\rm c}$ ($\gg {\rm g}\mu_{\rm B}B$)
is the relevant energy gap for the excitation process
in the QHE breakdown.
Within the QHE plateaus of $\nu$ = 3 and 5,
the shift  of the $V_{xx}$-$I$ curves ($\Delta I$) 
at $V_{xx}$ = 2~mV increases monotonically
with increasing the filling factor of Landau subbands
as shown in Figs.~\ref{fielddep}(a) and \ref{fielddep}(b),
i.e. the hysteresis is more prominent
when the Fermi energy locates closer to the upper Landau subband.

In a breakdown regime of QHE ($I > I_{\rm c}$),
current flows mainly in the inner bulk region of the 2DES.
However, in the Hall-bar geometry, 
edge channel transport may contribute to the DNP.
To know whether the bulk region is polarized, 
we studied another device with Corbino geometry,
where the edge-channel transport is completely absent.
We observed the similar shift of $V_{xx}$-$I$ curves
in breakdown regimes of odd-integer
QHE ($\nu$ = 1, 3, and 5) and detected the NMR signals.
These results definitely show that the nuclear spins
in the bulk region of the 2DES are polarized and detected
in this technique\cite{spatialdistribution}.
Details of the Corbino geometry experiment 
will be described elsewhere.

To summarize, we have demonstrated a  method for
electrical polarization of  nuclear spins 
in the inner bulk region of a quantum Hall conductor.
In a breakdown regime of odd-integer QHE,
the excitation of electrons to the upper Landau
subband  with opposite spin polarity
polarizes nuclear spins through the hyperfine interaction.
The dynamic nuclear polarization in turn
reduces the  spin-splitting energy of Landau subbands,
accelerating the QHE breakdown.

This work is supported by PRESTO, JST Agency, the Grant-in-Aid from 
MEXT (No. 17244120), and the Special Coordination Funds for Promoting 
Science and Technology.


\begin{thebibliography}{99}
\bibitem{energyconservation}
	Energy difference of electrons  between the initial
	and  final states in the ZT and IE excitations can be as small as
	the nuclear spin excitation energy.
\bibitem{wafer}
	In  the heterostructure used in the present study,
	the value of $I_{\rm c}$ is proportional to the
	width of conduction channel.
\bibitem{negativeI}
	When a current with opposite polarity $I$ = $-$2.0~$\mu$A is applied,
	$|V_{xx}|$ increases slowly as well.
\bibitem{spatialdistribution}
	Because the QHE breakdown develops along the current direction,
	the DNP may have spatial distribution along the channel.
\end{thebibliography}
\end{document}